\documentclass[11pt,reprint,english,aps,prc,showpacs,letter,longbibliography]{revtex4-1}

\usepackage[     bookmarks,
                 bookmarksopen = true,
                 bookmarksnumbered = true,
                 linktocpage,
                 colorlinks = true,
                 linkcolor = blue,
                 urlcolor  = blue,
                 citecolor = blue,
                 anchorcolor = green,
                 hyperindex = true,
                 hyperfigures]
        {hyperref}

\usepackage{graphicx}
\usepackage{dcolumn}
\usepackage{bm}
\usepackage{float}
\usepackage{amsmath}

\usepackage{color}
\usepackage{graphicx}
\definecolor{dgreen}{cmyk}{1.,0.,1.,0.2}        
\definecolor{orange}{cmyk}{0.,0.353,1.,0.}    

\newcommand\sect[1]{\section{#1}}

\begin{document}

\title{Probing Neutron Skin through Event-by-Event Pion Asymmetry in Heavy-ion collisions}

\author{Xu-Hua Tian}
\affiliation{Key Laboratory of Quark and Lepton Physics (MOE) \& Institute of Particle Physics, Central China Normal University, Wuhan 430079, China}

\author{Long-Gang Pang}
\email[]{lgpang@ccnu.edu.cn}
\affiliation{Key Laboratory of Quark and Lepton Physics (MOE) \& Institute of Particle Physics, Central China Normal University, Wuhan 430079, China}

\date{\today}%

\begin{abstract}

In this work, we propose a novel approach for probing the neutron skin thickness of gold (Au) by analyzing the event-by-event distribution of $\pi^{-}$ and $\pi^{+}$ yield differences. This is achieved through SMASH simulations of ultra-peripheral Au+Au collisions at $\sqrt{s_{\rm NN}}=3$ GeV. Our results demonstrate that the mean value of $\Delta n_{\pi} = n_{\pi^{-}} - n_{\pi^{+}}$, along with the Pearson correlation and mutual information between $(\pi^{-}+\pi^{+})$ and $(\pi^{-}-\pi^{+})$, all scale linearly with the neutron skin thickness. Moreover, the slope of the line connecting two distinct $\Delta n_{\pi}$ values in the event-by-event distribution also exhibits a linear dependence on the neutron skin thickness. The most sensitive $\Delta n_{\pi}$ pairs are identified as $(-1, 1)$, $(-1, 2)$, $(0, 1)$, and $(0, 2)$. These findings establish a new pathway for determining the neutron skin thickness. Finally, by comparing SMASH and UrQMD simulations under identical initial conditions, we observe that individual slope values depend on the specific collision model. However, by extracting slopes from multiple $\Delta n_{\pi}$ pairs in experimental event-by-event data and inferring the corresponding neutron skin thickness, one can assess which model better aligns with the true physical value.

\end{abstract}
\maketitle

\sect{Introduction}

The neutron skin thickness, defined as the difference between the root-mean-square (rms) radii of neutron and proton density distributions, $\Delta r_{\text{np}} = \langle r_{\text{n}}^2 \rangle^{1/2} - \langle r_{\text{p}}^2 \rangle^{1/2}$ \cite{Brown:2000pd}, serves as a fundamental probe of the structure of atomic nuclei. This parameter is extremely important for constraining the equation of state of asymmetric nuclear matter \cite{LiLi:2022kvc,Centelles:2008vu,Li_2017,Reinhard:2016mdi,Typel:2001lcw,Chen:2005ti,Roca-Maza:2011qcr,Chen:2010qx,Huang:2025uzc},  which in turn governs the structure \cite{PhysRevLett.86.5647,Horowitz:2000xj,Huang:2025uzc} and evolution of neutron stars \cite{Brown:2000pd,PhysRevLett.127.192701,Fattoyev:2017jql,Ji:2019hxe}. While the proton radius can be accurately measured using experimental techniques such as electron and proton scattering, the precise determination of the neutron distribution radius remains a significant challenge due to the charge-neutral nature of the neutron. 

Experimentally, the neutron skin thickness can be measured using parity-violating electron scattering experiments \cite{PREX:2021umo,Abrahamyan:2012gp}. 
The underlying principle is as follows: When longitudinally polarized electrons are incident on an atomic nucleus, the weak interaction induces a difference between the scattering cross-sections of left-handed and right-handed electrons, giving rise to a parity-violating asymmetry ($A_{PV}$). The weak charge form factor, $F_w(Q^2)$, can be extracted from the experimentally measured $A_{PV}$. Subsequently, the weak charge radius, $R_w$, is obtained by fitting the experimentally determined $F_w(Q^2)$ with theoretical models. Within various nuclear structure models, $R_w$ exhibits a strong linear correlation with the neutron skin thickness. Ultimately, a strong linear relationship was found between the $A_{PV}$ and the neutron skin thickness, allowing the $A_{PV}$ to be used to determine the neutron skin thickness. A key advantage of this method is its insensitivity to strong-interaction uncertainties. However, the parity-violating asymmetry from weak interactions is exceedingly small (on the order of ppm), requiring large datasets and high luminosity for statistically significant measurements. The Jefferson Laboratory has applied this technology to measure the neutron skin thickness of lead and calcium-48, with the results being \(0.283 \pm 0.071~\text{fm}\) and \(0.121 \pm 0.026\ (\text{exp}) \pm 0.024\ (\text{model})\), respectively\cite{CREX:2022kgg,PREX:2021umo}, yet the resulting neutron skins yield contradictory predictions for the slope of the symmetry energy. Currently, no single theoretical model can reconcile the CREX and PREX-2 experimental data \cite{Tagami:2022spb,Reinhard:2022inh,Yuksel:2022umn,Zhang:2022bni}.

Relativistic isobar collisions can also be used to determine the neutron skin thickness. Final-state observables in relativistic heavy-ion collisions exhibit strong sensitivity to the initial nuclear structure. By treating the neutron skin as a free parameter, one can perform collision simulations using theoretical models; combining these simulations with experimental data from the Large Hadron Collider enables the extraction of the neutron-skin thickness. Specifically, observables such as charged-particle multiplicity \cite{Li:2019kkh}, average transverse momentum \cite{Xu:2021uar}, Neutron-Proton Yield Ratio \cite{Yang:2023bwm}, spectator neutrons \cite{Zhang:2025voj}, and — more directly — electromagnetic charge multiplicity, which distinguishes proton and neutron distributions \cite{Xu:2021qjw}, can be employed to constrain the neutron-skin thickness. Additionally, Bayesian analyses incorporating multiple experimental observables can be used to infer the neutron skin; however, the choice of observables in the fitting procedure influences the resulting neutron-skin thickness, leading to variations depending on the selected dataset \cite{Giacalone:2023cet}. Currently, most studies on neutron skin thickness focus on lead, with only a limited number addressing gold. However, gold (Au) has abundant experimental collision data \cite{STAR:2008med,STAR:2004jwm,STAR:2013qio,1993Nuclear}. Therefore, investigating the neutron skin thickness of gold nuclei not only enhances our understanding of their structure and helps derive the nuclear equation of state, but also facilitates comparisons with experimental results. 

The STAR Collaboration has introduced a nuclear tomography method based on quantum interference to determine the strong-interaction nuclear radius, by examining the angular distribution of decay products from vector mesons generated in polarized photon-nucleus collisions. The systematic discrepancy between this radius and the charge radius is ascribed to the neutron skin effect, thereby facilitating indirect measurements of the neutron skin thickness in Au. This study reports a neutron skin thickness of 0.17 fm for gold\cite{STAR:2022wfe}. Theoretically, existing studies have employed the Hartree–Fock–Bogoliubov (HFB) theory and the Relativistic Continuum-Hartree–Bogoliubov (RCHB) model to calculate the neutron skin thickness of gold isotopes \cite{AV2024122889}.

Transport models can also be employed to predict neutron skin thickness by simulating pion production in heavy-ion collisions \cite{Tellez-Arenas:1987tls,Lombard:1988pj}. It has been proposed that the $\pi^{-}/\pi^{+}$ yield ratio may constrain the neutron skin thickness \cite{Wei:2013sfa}. In high-energy heavy-ion collisions, the $\pi^{-}/\pi^{+}$ ratio is sensitive to the $n/p$ ratio in high-density nuclear matter, which reflects the nuclear symmetry energy at high densities \cite{Li:2002qx}. Given that neutron skin thickness is also a sensitive probe of the symmetry energy \cite{Brown:2000pd,Typel:2001lcw,PhysRevLett.86.5647,Li:2019kkh}, the $\pi^{-}/\pi^{+}$ yield ratio can potentially constrain it. Studies indicate that in non-central collisions, the neutron skin effect significantly influences the $\pi^{-}/\pi^{+}$ ratio \cite{Wei:2013sfa,Hartnack:2019tlu}. In these studies, the pion yield is averaged over multiple events. Furthermore, different transport models exhibit substantial discrepancies in predicting this ratio, indicating strong model dependence \cite{TMEP:2023ifw,Hong:2013yva,Zhang:2025gro}.

In contrast, our study focuses on event-by-event analysis of the difference between  $\pi^{-}$ and $\pi^{+}$ yields and introduces a new observable: $\pi^{-} - \pi^{+}$ instead of $\pi^{-}/\pi^{+}$. This approach circumvents issues arising from zero $\pi^{+}$ counts in individual events, thereby enabling more accurate constraints on neutron skin thickness. Although our method retains some model dependence, the event-wise analysis produces multiple instances of $\Delta n_{\pi}$. By comparing the slopes of multiple $\Delta n_{\pi}$ datasets with experimental results, this model dependence can be effectively reduced.

\begin{figure}[htp]
\centering
\includegraphics[width=0.48\textwidth]{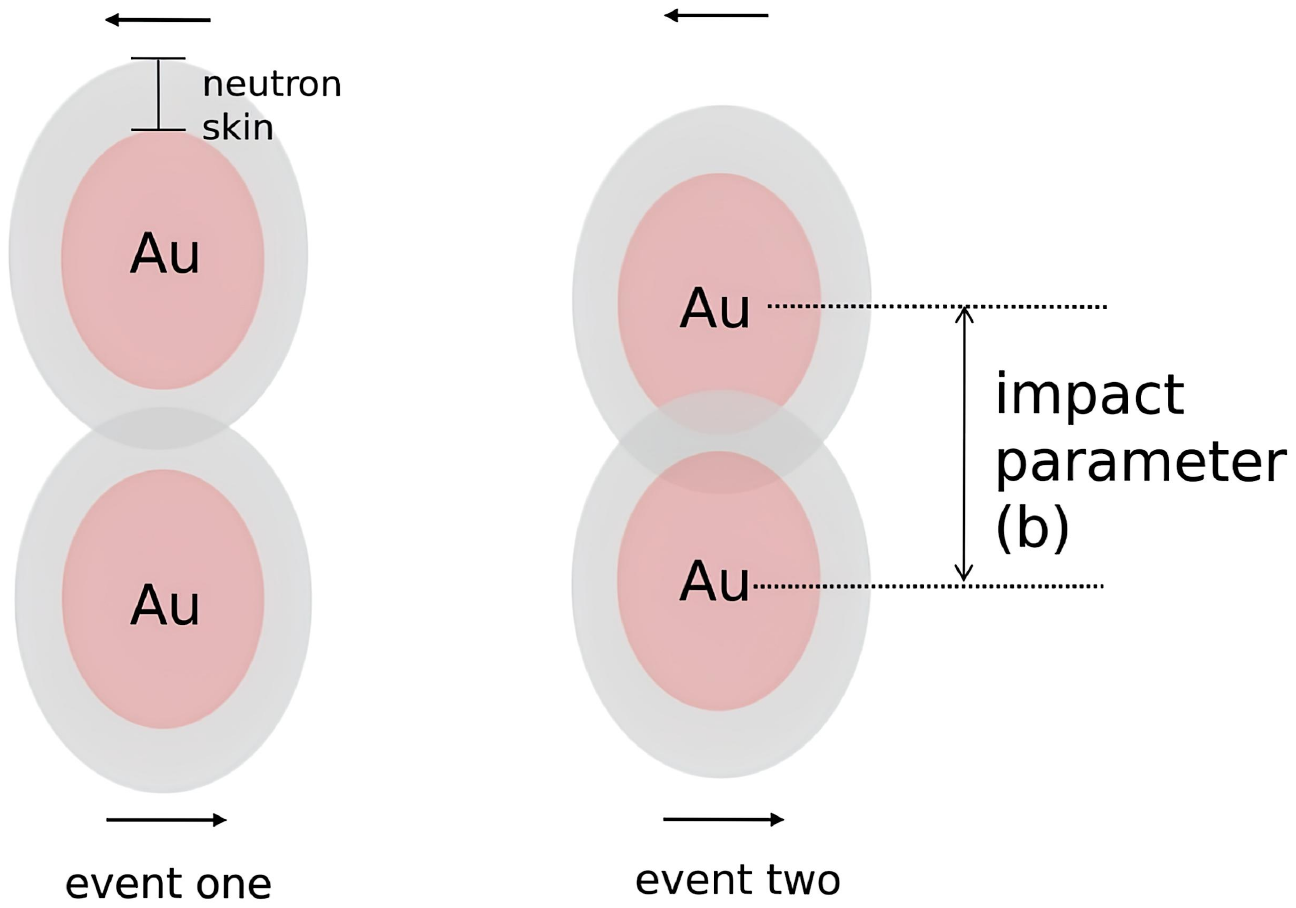}
\caption{(Color online) 
The neutrons and protons are mixed in the nucleus, and the root-mean-square radius of the neutron distribution is larger than that of the proton distribution, forming the neutron skin structure shown in the gray area.\label{fig:off-central collision}}
\end{figure}

Figure \ref{fig:off-central collision} illustrates a schematic representation of Au–Au peripheral collisions with varying overlap regions. The red core region contains both protons and neutrons, with neutrons exhibiting higher density. The gray outer region of the Au nucleus represents the neutron skin. The transverse separation between the centers of the two colliding Au nuclei in the plane perpendicular to the beam axis is defined as the impact parameter b, which varies across different collision events. As depicted, Event 1 features only neutron–neutron (nn) collisions at larger impact parameters, while Event 2 begins to exhibit proton–neutron (np) collisions as the impact parameter decreases. Consequently, the event-by-event distribution of impact parameters encodes information about neutron skin.

\sect{Method}
In this study, we employed the SMASH (Simulating Many Accelerated Strongly-interacting Hadrons) dynamical evolution model to simulate the outcomes of Au-Au peripheral collisions under various neutron skin thickness conditions. The SMASH model is an advanced hadronic transport approach that effectively describes the final stages of strong interactions in low-energy heavy-ion collisions, particularly suitable for characterizing strongly interacting systems in non-equilibrium states. This model provides an effective solution to the relativistic Boltzmann equation \cite{Sciarra:2024gcz} and has been widely applied in various studies on particle collisions.

In SMASH, we sample the spatial coordinates of 118 neutrons and 79 protons in the colliding Au nucleus using a 3-parameter Woods-Saxon distribution $\rho(r)$, 
\begin{equation}
p(r) =  4\pi r^2 \rho(r) = \frac{4\pi r^2\rho_0}{\exp\left(\frac{r - r_0}{d}\right) + 1} 
\end{equation}
where $\rho_0 \sim 0.16\ {\rm fm}^3$ is the saturation density, $r_0$ is the half-density radius of the nucleus, $d$ is the surface diffusion parameter. The coefficient $4\pi \rho_0$ do not affect the Monte Carlo sampling. In practice, we sample 79 protons and 118 neutrons from this distribution, whose effective $\rho_0$ would be slightly different from $ 0.16\ {\rm fm}^3$.

The neutron skin thickness distribution manifests in two forms: skin-type and halo-type. Low-energy nuclear experiments suggest that most atomic nuclei exhibit a halo-type neutron skin \cite{Trzcinska:2001sy}. In this work, we adopt the halo-type configuration, wherein the initial sampling of protons and neutrons assumes equal half-density radii \( r_0 = 6.638 \) fm for both species, while the neutron density diffusion parameter \( d_n \) exceeds that of protons \( d_p \). The neutron skin thickness is modulated by varying \( d_n \). For protons, the surface diffusion parameter is fixed at \( d_p = 0.535 \) fm. For neutrons, we introduce a parameter \( c \) in the relation \( d_n = d_p + c \) within the Woods-Saxon distribution to adjust the neutron skin thickness \( \Delta r_{np} \). 

\begin{figure}[htp]
\centering
\includegraphics[width=0.48\textwidth]{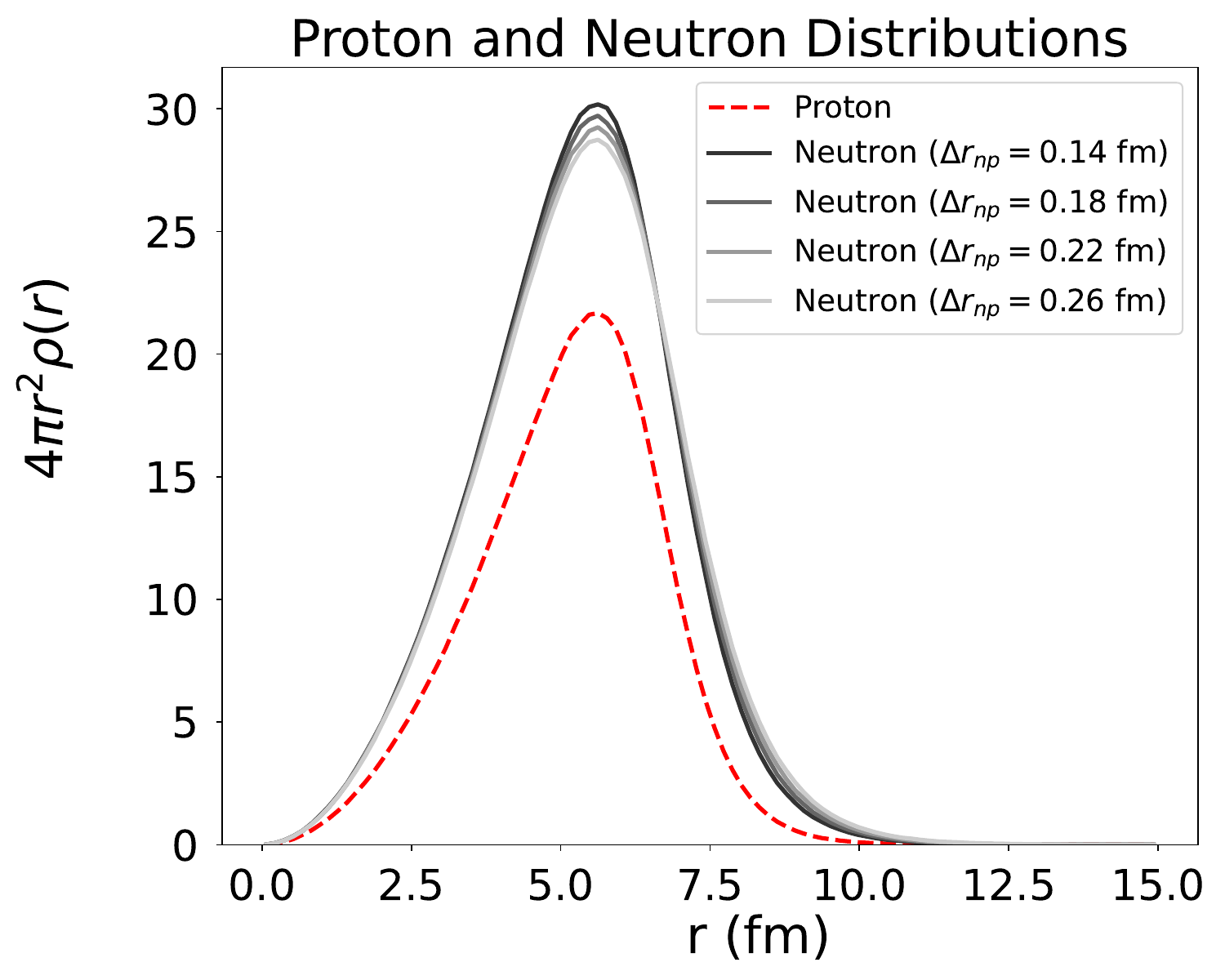}
\caption{(Color online) The sampling results of the initial distribution are shown in the figure, where the red dashed line represents the proton sampling results, and the dark gray to light gray solid lines correspond to the neutron sampling results for neutron skin thicknesses of 0.14 fm, 0.18 fm, 0.22 fm, and 0.26 fm, respectively. \label{fig:n_p_sampling}}
\end{figure}

We computed \( \Delta r_{np} \) for a range of \( d_n \) values, enabling the determination of \( c = d_n - d_p \) corresponding to neutron skin thicknesses of \( \Delta r_{np} = 0.14, 0.16, 0.18, 0.20, 0.22, 0.24, \) and \( 0.26 \) fm via linear interpolation. Figure \ref{fig:n_p_sampling} illustrates the initial proton and neutron distributions for four selected neutron skin thickness values.

We propose a novel observable, the distribution of $n_{\pi^{-}} - n_{\pi^{+}}$, to constrain the neutron skin thickness. Here, the $n_{\pi^{-}}$ denotes the number of $\pi^{-}$ produced in a single event of Au+Au collisions, and $n_{\pi^{+}}$ represents the number of $\pi^{+}$ in a single event. Traditionally, it has been observed that the yield ratio $\langle n_{\pi^{-}} \rangle / \langle n_{\pi^{+}} \rangle$ in peripheral heavy ion collisions is sensitive to the neutron skin thickness, where $\langle n_{\pi^{\pm}} \rangle$ represents the mean values of $\pi^{\pm}$ produced in heavy ion collisions. We hypothesize that the event-by-event distribution of $n_{\pi^{-}} / n_{\pi^{+}}$ might be sensitive to the neutron skin thickness. However, in ultra-peripheral collisions where the dominant process is neutron-neutron collisions, the $\pi^+$ yield can be 0 in a single collision. Even if we consider $n_{\pi^{+}} / n_{\pi^{-}}$ instead, the observable can also be ill-defined, because of event-by-event fluctuations. To address this issue, we utilize the event-by-event distribution of $n_{\pi^{-}} - n_{\pi^{+}}$.

We utilize the 20\% most peripheral collisions to study the sensitivity of $n_{\pi^{-}} - n_{\pi^{+}}$ distributions to neutron-skin thicknesses. 
To efficiently accumulate a substantial dataset for peripheral collisions within the 80\%-100\% centrality range, we implemented the following methodology: First, Au+Au collision simulations were performed with impact parameters ranging from 0 fm to 16 fm. From these simulations, events corresponding to the lowest 20\% of particle multiplicities were selected, and the minimum impact parameter $b_{\rm min}$ and the maximum particle multiplicity $N_{\rm max}$ for these events were recorded. Subsequently, a second set of Au+Au collision simulations was conducted with impact parameters between $b_{\rm min}$ and 16 fm, retaining only those events with particle multiplicities below $N_{\rm max}$. This approach enabled the acquisition of a sufficient number of off-central collision events for analysis.

To verify the model dependence of the $n_{\pi^-} - n_{\pi^+}$ distribution, we also used the UrQMD model to simulate the results of ultra-peripheral Au-Au collisions at $\sqrt{s_{NN}}=3$ GeV. For comparative studies, the UrQMD model and the SMASH model adopted the same neutron skin thickness settings and collision parameter settings. Due to the limitation of computational resources, the Coulomb force was not included in the SMASH model, while the UrQMD model incorporated the effect of the Coulomb force. Moreover, due to the limitation of computational speed, the UrQMD model only performed simulations for three neutron skin thicknesses: 0.16 fm, 0.20 fm, and 0.24 fm.

\sect{Results}

\begin{figure}[!htp]
\centering
\includegraphics[width=0.48\textwidth]{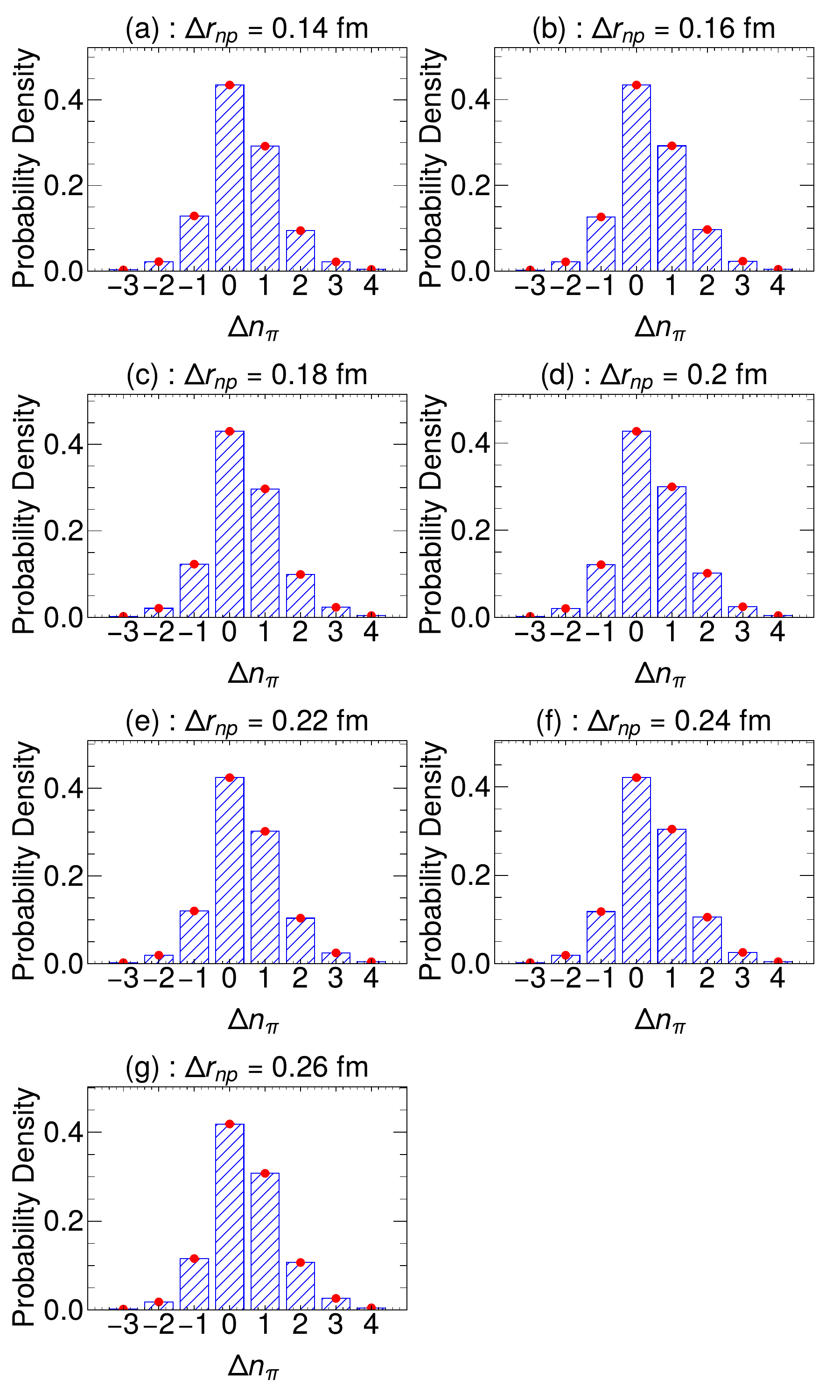}
\caption{(Color online) Density plot of $\Delta n_{\pi}$ from Au-Au collisions at various neutron skin thicknesses. The red part indicates the relative error.\label{fig:ebe_delta_pi}}
\end{figure}

Fig.~\ref{fig:ebe_delta_pi} illustrates the normalized event-by-event distribution of the difference between the number of $\pi^-$ and $\pi^+$ mesons, denoted as $\Delta n_{\pi}$ (defined as \(\Delta n_{\pi} = n_{\pi^-} - n_{\pi^+})\), for each event under varying neutron skin thickness conditions in the 20\% most peripheral Au+Au collisions at $\sqrt{s_{NN}}=3$ GeV. Subfigures (a) to (g) in Fig.~\ref{fig:ebe_delta_pi} correspond to neutron skin thicknesses of 0.14 fm, 0.16 fm, 0.18 fm, 0.20 fm, 0.22 fm, 0.24 fm, and 0.26 fm, respectively. In Fig.\ref{fig:ebe_delta_pi}, the range of \(\Delta n_{\pi}\) is set from -3 to 4, where \(\Delta n_{\pi} < 0\) indicates that the number of \(\pi^-\) mesons is fewer than the number of \(\pi^+\) mesons in the event, and \(\Delta n_{\pi} > 0\) indicates that the number of \(\pi^+\) mesons is fewer than the number of \(\pi^-\) mesons. All subfigures reveal an asymmetric distribution about \(\Delta n_{\pi}=0\), with the distribution systematically shifted toward positive values. This demonstrates that \(\pi^-\) production consistently exceeds \(\pi^+\) production across all examined neutron skin configurations.

Further analysis indicates a systematic correlation between the specific \(\Delta n_{\pi}\) values and the neutron skin thickness \(\Delta r_{np}\). For instance, at \(\Delta r_{np} =  0.14\) fm, the number of events with \(\Delta n_{\pi} = -1\) significantly exhibits a higher occurrence frequency than those with \(\Delta n_{\pi} = 2\). As the neutron skin thickness progressively increases to 0.16 fm, 0.18 fm, 0.20 fm and 0.22 fm, the gap between these two values progressively diminishes. As the neutron skin thickness further increases to 0.24 fm and 0.26 fm, the two values become almost equal. Similar correlations are also observed for other \(\Delta n_{\pi}\) values. 

\begin{figure}[!htp]
\centering
\includegraphics[width=0.48\textwidth]{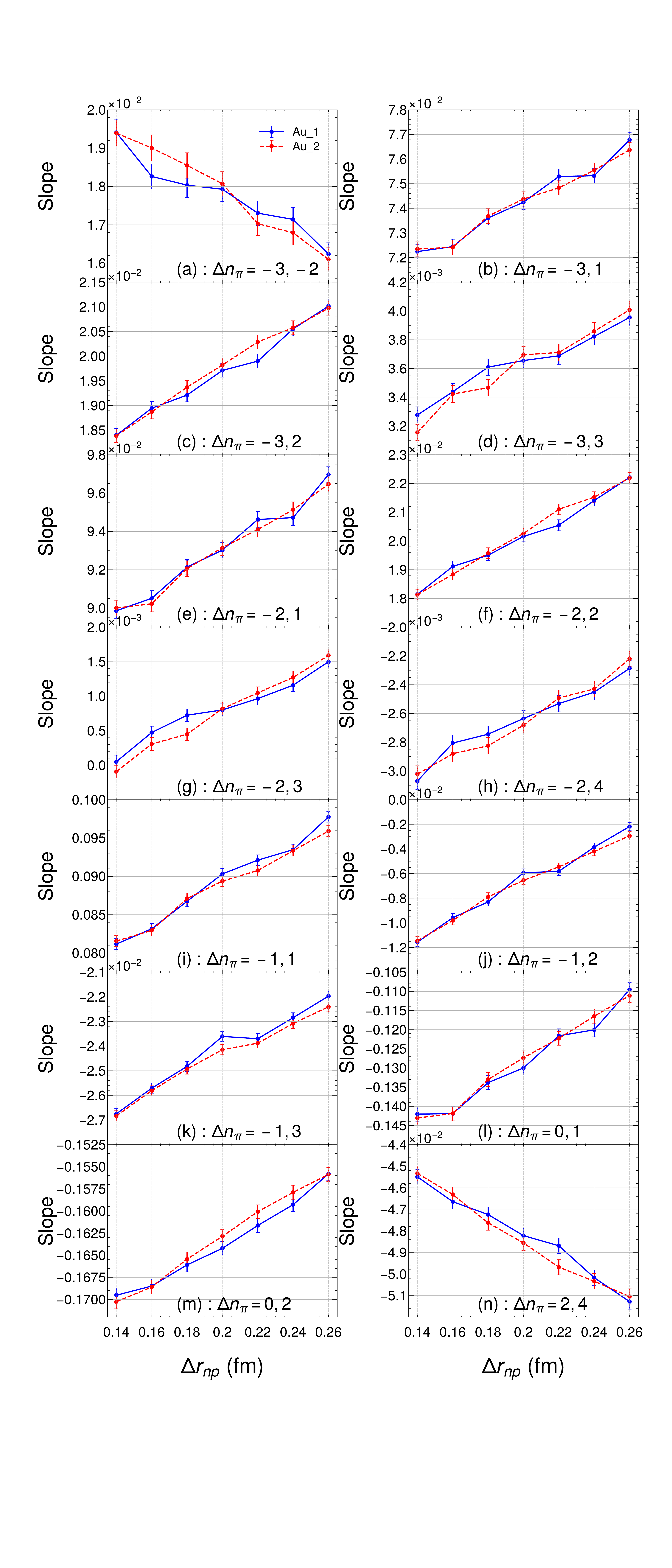}
\caption{(Color online) The relationship between the slope formed by two different $\Delta n_{\pi}$ values in the histogram and the neutron skin thickness is shown. The two lines in the figure represent two sets of SMASH collision data with the same initial state, and the short-line segments denote the relative uncertainties.\label{fig:slope_vs_skin_thickness_subplot}}
\end{figure}

Fig.~\ref{fig:slope_vs_skin_thickness_subplot} illustrates the correlation between the slopes of different \(\Delta n_{\pi}\) pairs and neutron skin thickness more intuitively, in the event-by-event distributions of \(\Delta n_{\pi}\) shown in Fig.~\ref{fig:ebe_delta_pi}. 
The slope parameter $k$ is defined as the slope of the line intersecting the event-by-event distribution for two distinct $\Delta n_{\pi}=n_{\pi^{-}} - n_{\pi^{+}}$ values, shown below
\begin{align}
    k = {1 \over a} \left[P(\Delta n_{\pi2}) - P(\Delta n_{\pi1}) \right] 
\end{align}
where $a = \Delta n_{\pi 1} - \Delta n_{\pi 2}$,
\(\Delta n_{\pi1}\) and \(\Delta n_{\pi2}\) represent two different \(\Delta n_{\pi}\) values.
Subfigures (a) to (n) in Fig.\ref{fig:slope_vs_skin_thickness_subplot} systematically display the relationships between neutron skin thickness and the slope pairs, e.g., \(\Delta n_{\pi} = (-3, -2)\).
To verify the effect of statistical error of the $n_{\pi^-} - n_{\pi^+}$ distribution, we performed two SMASH collision simulations with identical initial conditions: each run contained $600\,000$ events with impact parameters from $b_{\mathrm{min}}$ to 16 fm. After rejecting events whose particle multiplicities exceeded $N_{\max}$ that gives the boundary of $20\%$ most peripheral collisions, both runs retained approximately $210\,000$ events. As can be seen from Fig.\ref{fig:slope_vs_skin_thickness_subplot}, these two sets of data are almost identical within the error range.

As shown in Fig.\ref{fig:slope_vs_skin_thickness_subplot}, the most significant variations occur for cases (i) (\(\Delta n_{\pi} = -1, 1\)), (j) (\(\Delta n_{\pi} = -1, 2\)), (l) (\(\Delta n_{\pi} = 0, 1\)), and (m) (\(\Delta n_{\pi} = 0, 2\)), demonstrating their strong sensitivity to neutron skin thickness. 
In cases (i), (j), (l) and (m), the slope exhibits an approximately linear increase with increasing neutron skin thickness. This linear dependence enables the determination of neutron skin thickness from experimentally measured slopes between different \(\Delta n_{\pi}\) pairs. 
For instance, if a measured slope of 0.090 is obtained for \(\Delta n_{\pi} = -1, 1\) in case (j), our calculations suggest a corresponding neutron skin thickness of 0.20 fm.
While this individual determination may be subject to model uncertainties, consistency across multiple \(\Delta n_{\pi}\) pairs would significantly enhance confidence in the derived \(\Delta r_{np} = 0.20\) fm value.

The sub-figures in Fig.\ref{fig:slope_vs_skin_thickness_subplot} can be categorized into two distinct groups. In the first group (panels a, n), the slopes decrease with increasing neutron skin thickness, while in the second group (panels b, c, d, e, f, g, h, i, j, k, l, m), the slopes exhibit an opposite trend.
Upon examining the \(\Delta n_{\pi}\) for these two categories, we identified a clear pattern governing the slope behavior. For the first group, both values of the \(\Delta n_{\pi}\) pairs share the same sign, whereas for the second group, they possess opposite signs. This observation suggests that the mean of the \(\Delta n_{\pi}\) distribution shifts toward more positive values as the neutron skin thickness increases.
This finding underscores the rationale behind computing the slopes of different \(\Delta n_{\pi}\) pairs across varying neutron skin thicknesses.

The error bars in Fig.\ref{fig:slope_vs_skin_thickness_subplot} are calculated using the error propagation method for the slope parameter $k$, 
\begin{equation}
    \Delta k = {1\over |a|}\sqrt{\delta_1^2 + \delta_2^2}
\end{equation}
where \(\delta_1\) and \(\delta_2\) are the corresponding errors of \(P(\Delta n_{\pi})\) for  \(\Delta n_{\pi1}\) and \(\Delta n_{\pi2}\), whose absolute error is computed using $\delta = P(\Delta n_{\pi})/\sqrt{N}$.

\begin{figure}[!htp]
\centering
\includegraphics[width=0.48\textwidth]{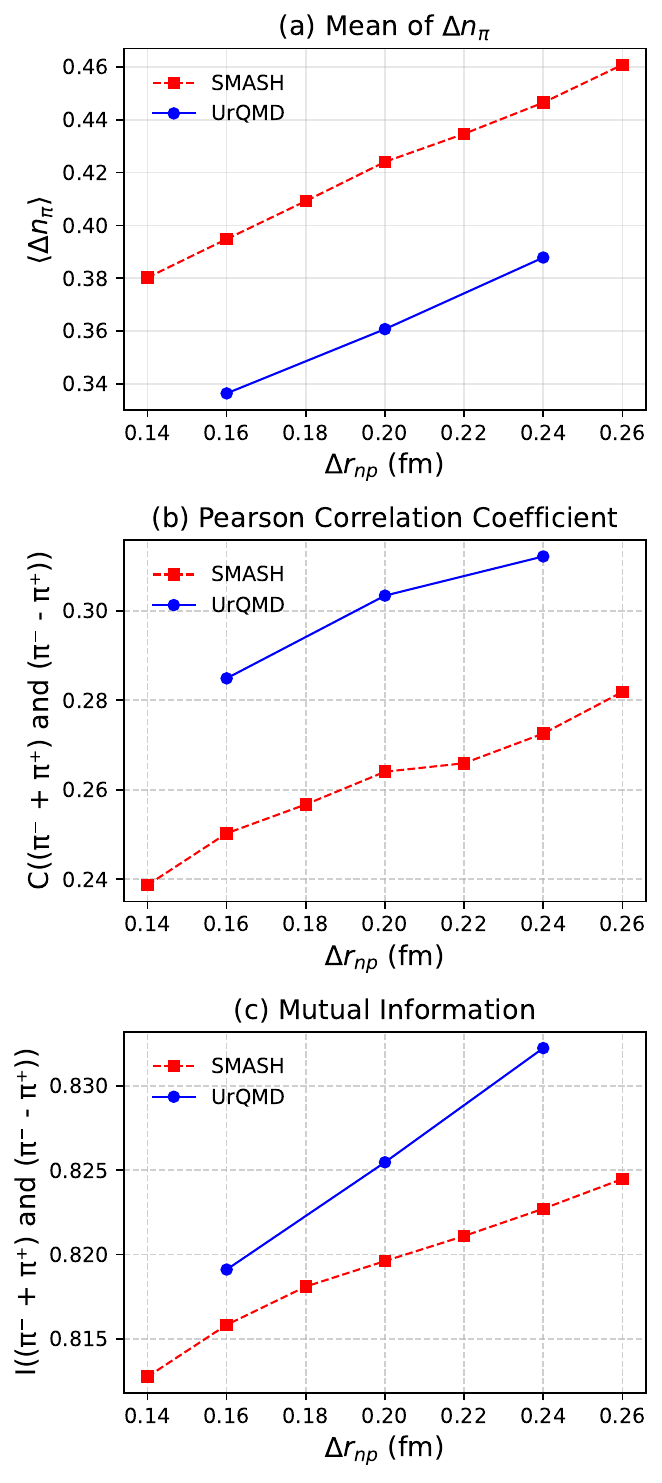}
\caption{(Color online) Comparison of the relationship between the neutron skin thickness $\Delta r_{np}$ and the following three physical quantities: (a) the mean value of $\Delta n_{\pi}$; (b) the Pearson correlation between $(\pi^{-}+\pi^{+})$ and $(\pi^{-}-\pi^{+})$; (c) the mutual information between $(\pi^{-}+\pi^{+})$ and $(\pi^{-}-\pi^{+})$.  
The red dashed line represents the results from the SMASH model, while the blue solid line represents the results from the UrQMD model.\label{fig:three_subplots}}
\end{figure}

Fig.\ref{fig:three_subplots} (a) presents the variation of the mean value of $\Delta n_{\pi}$ with the neutron skin thickness, obtained from collision simulations using the SMASH model and the UrQMD model, respectively. The red dashed line represents the results from the SMASH model, and the blue solid line represents the results from the UrQMD model. Observing Fig.\ref{fig:three_subplots} (a), it is found that the mean value of \(\Delta n_{\pi}\) increases with the increase of neutron skin thickness. The results of Figure \ref{fig:three_subplots} (a) are consistent with the above inference. In Fig.\ref{fig:three_subplots} (a), the mean value of $\Delta n_{\pi}$ exhibits a nearly linear relationship with the neutron-skin thickness; hence, combining the theoretically predicted curve with the experimentally measured mean $\Delta n_{\pi}$ can be used to infer the neutron-skin thickness. However, predictions based solely on the mean carry a significant model dependence and do not effectively reduce this uncertainty. Employing multiple $\Delta n_{\pi}$ pairs to predict the neutron-skin thickness can substantially mitigate such model dependence.

Fig.\ref{fig:three_subplots} (b) presents the Pearson correlation between $(\pi^{-} + \pi^{+})$ and $(\pi^{-} - \pi^{+})$ as a function of neutron-skin thickness.  The red dashed and blue solid lines represent the results from the SMASH and UrQMD models, respectively. A positive correlation is observed between the Pearson coefficient and the neutron-skin thickness, indicating that thicker neutron skins correspond to stronger linear correlations between $(\pi^{-} + \pi^{+})$ and $(\pi^{-} - \pi^{+})$. For any fixed skin thickness, the UrQMD model yields higher values than SMASH, suggesting that UrQMD predicts a stronger linear correlation. However, the Pearson correlation exhibits significant model dependence, which precludes the extraction of neutron-skin thickness based solely on this observable.

Fig.\ref{fig:three_subplots} (c)  shows the mutual information between $(\pi^{-} + \pi^{+})$ and $(\pi^{-} - \pi^{+})$ as a function of neutron-skin thickness. Mutual information, which captures nonlinear correlations more effectively than Pearson correlation, is shown by the red dashed line for the SMASH model and the blue solid line for the UrQMD model. The mutual information trend resembles that of the Pearson correlation, though the two correlation curves are less parallel in comparison. The models show reasonable agreement for thin neutron skins, while for thick skins, UrQMD yields a stronger amplification of the correlation than SMASH.

\begin{figure}[!htp]
\centering
\includegraphics[width=0.48\textwidth]{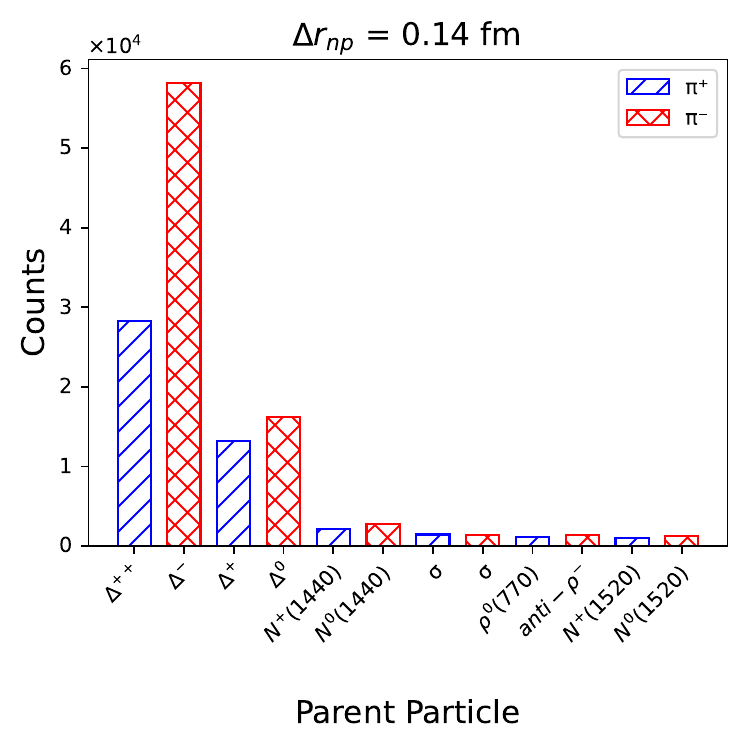}
\caption{(Color online) A histogram of the six most abundant parent particles producing $\pi^-$ and $\pi^+$ mesons via resonant decay in Au collisions at a center-of-mass energy of $\sqrt{s_{NN}} = 3$ GeV with a neutron skin thickness of 0.14 fm.\label{fig:decay_particle}}
\end{figure}

Fig.\ref{fig:decay_particle} presents a statistical analysis of the production mechanisms of \(\pi^-\) and \(\pi^+\) mesons.
The plot illustrates the six most abundant parent particles contributing to  $\pi^-$ and $\pi^+$ production via resonant decay in SMASH simulations of Au+Au ultra-peripheral collisions at $\sqrt{s_{NN}} = 3$ GeV, assuming a neutron skin thickness of 0.14 fm for the Au nucleus. 
Our analysis indicates that \(\pi^-\) and \(\pi^+\) mesons are predominantly generated through resonance decays, accounting for approximately 96\% of their total yield. In the figure, blue hatched bars denote $\pi^{+}$ mesons, and red grid-patterned bars denote $\pi^{-}$ mesons. It is observed that \(\pi\) mesons are primarily produced by the resonant state \(\Delta\) decay.

The decays of $\Delta$ resonance states into pions are shown below \cite{Bass:1993ce}:
 \begin{equation}
|\Delta^{++}\rangle \leftrightarrow |p\rangle|\pi^+\rangle
\end{equation}
 \begin{equation}
|\Delta^{+}\rangle \leftrightarrow \sqrt{\frac{2}{3}}|p\rangle |\pi^0\rangle + \sqrt{\frac{1}{3}}|n\rangle |\pi^+\rangle
\end{equation}
 \begin{equation}
|\Delta^{0}\rangle \leftrightarrow \sqrt{\frac{2}{3}}|n\rangle |\pi^0\rangle + \sqrt{\frac{1}{3}}|p\rangle |\pi^-\rangle
\end{equation}
 \begin{equation}
\Delta^{-} \leftrightarrow |n\rangle |\pi^-\rangle
\end{equation}

It is evident that $\pi^{+}$ primarily originate from decays of $\Delta^{++}$ and $\Delta^{+}$, whereas $\pi^{-}$ mainly stem from decays of $\Delta^{0}$ and $\Delta^{-}$.
Table.\ref{tab:clebsch} summarizes the production channels of $\Delta$ resonances and the corresponding Clebsch-Gordan coefficients. The results indicate that $\Delta$ resonances are predominantly produced through inelastic nucleon-nucleon scatterings.
By combining the decay mechanism of $\Delta$ resonances with the production channels listed in Table~\ref{tab:clebsch}, 
we conclude that neutron-neutron (nn) collisions primarily produce  $\pi^-$, while proton-proton (pp) collisions predominantly yield $\pi^+$. Given that the Au nucleus contains more neutrons than protons, the number of nn collisions exceeds that of pp collisions on average, particularly in peripheral collisions. Consequently, more $\pi^{-}$ than $\pi^{+}$ are observed in Au+Au collisions, which also explains the rightward shift of the peak in Fig.\ref{fig:ebe_delta_pi}. 

\begin{table}[h]
\centering
\caption{Expected isospin and symmetry factors for number of reactions within isospin groups at equilibrium. The first numeric column is a Clebsch-Gordan factor, the second column is symmetry factor, the third one is their product\cite{SMASH:2016zqf}}

\begin{tabular}{|l|c|c|c|}
\hline
Reaction & Clebsch & Symmetry & Total \\ \hline
$pp \to p\Delta^{+}$      & 1/4 & 1/2 & 1/8 \\ 
$pp \to n\Delta^{++}$     & 3/4 & 1/2 & 3/8 \\  
$pn \to n\Delta^{+}$      & 1/4 & 1   & 2/8 \\ 
$pn \to p\Delta^{0}$      & 1/4 & 1   & 2/8 \\ 
$nn \to p\Delta^{-}$      & 3/4 & 1/2 & 3/8 \\ 
$nn \to n\Delta^{0}$      & 1/4 & 1/2 & 1/8 \\   
$pp \to \Delta^{0}\Delta^{++}$      & 6/20  & 1/2 & 18/120 \\ 
$pp \to \Delta^{+}\Delta^{+}$       & 8/20  & 1/4 & 12/120 \\ 
$pn \to \Delta^{-}\Delta^{++}$ & 67/120 & 1 & 67/120 \\ 
$pn \to \Delta^{+}\Delta^{0}$  & 43/120 & 1 & 43/120 \\ 
$nn \to \Delta^{+}\Delta^{-}$       & 6/20  & 1/2 & 18/120 \\ 
$nn \to \Delta^{0}\Delta^{0}$       & 8/20  & 1/4 & 12/120 \\ \hline
\end{tabular}
\label{tab:clebsch}
\end{table}

\begin{figure}[!htp]
\centering
\includegraphics[width=0.48\textwidth]{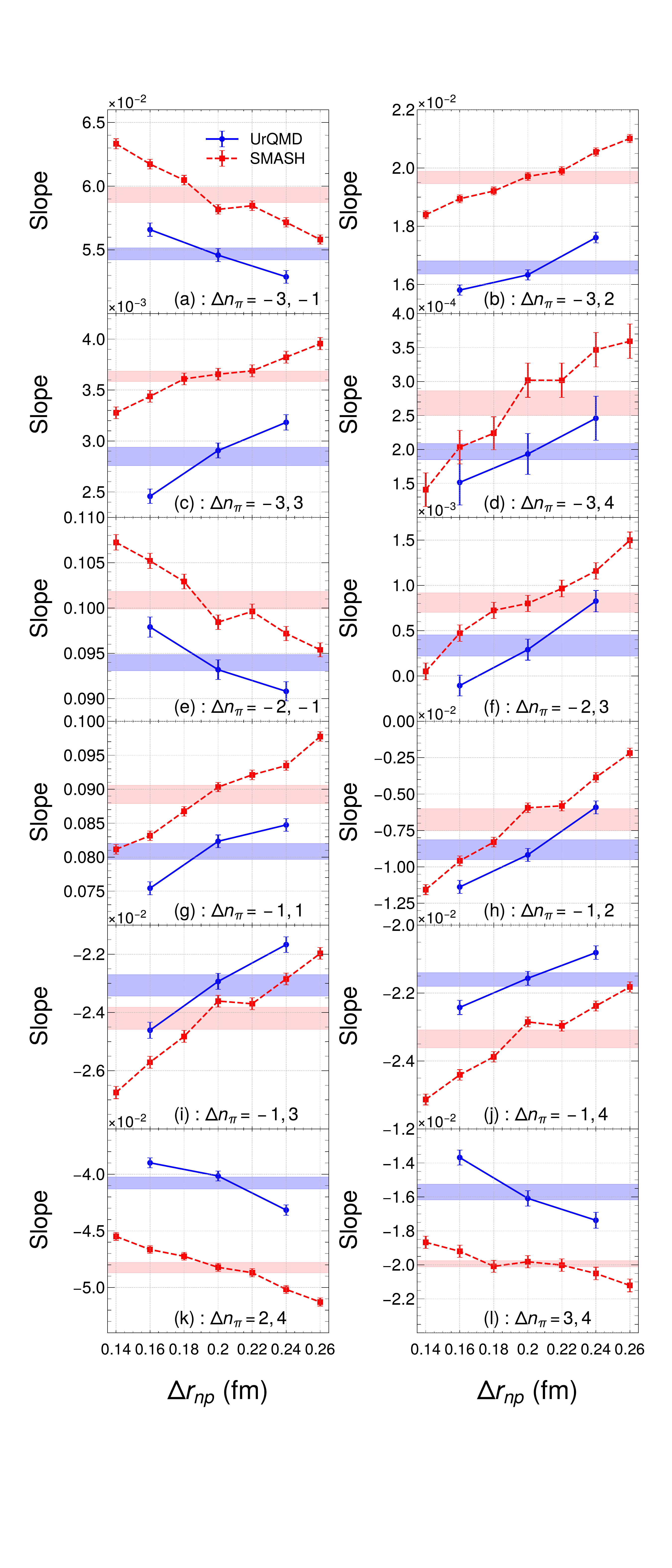}
    \caption{(Color online)The figure illustrates the correlation between the slope derived from two distinct $\Delta n_{\pi}$ values in the histograms of the SMASH and UrQMD models and the neutron skin thickness. The blue solid line denotes the UrQMD model result, whereas the red dashed line represents the SMASH model simulation. The blue band indicates the slope range corresponding to a neutron skin thickness of 0.19–0.21~fm in the UrQMD model data, while the red band shows the corresponding range for the SMASH model data.
\label{fig:slope_comparison_band}}
\end{figure}

Fig.\ref{fig:slope_comparison_band}  shows the relationship between neutron skin thickness and the slope for different $\Delta n_{\pi}$, as obtained from the SMASH (red dashed lines) and UrQMD (blue solid lines) models under identical initial conditions and statistical methods.
The blue and red bands represent the range of slope values corresponding to neutron skin thicknesses from 0.19 fm to 0.21 fm for the UrQMD and SMASH models, respectively.
Given the strong linear correlation between the slope parameter and neutron skin thickness for these $\Delta n_{\pi}$ pairs, a linear 
regression is employed to estimate the slopes at neutron skin thicknesses of 0.19 fm and 0.21 fm.
The SMASH model includes results for seven neutron skin thicknesses (0.14, 0.16, 0.18, 0.20, 0.22, 0.24, and 0.26 fm), whereas the UrQMD model provides results for three (0.16, 0.20, and 0.24 fm). For the same $\Delta n_{\pi}$ pair, the trends in the relationship  between neutron skin thickness and slope are similar between the SMASH and UrQMD models, but the neutron skin thickness corresponding to the same slope differs significantly, which demonstrates that the slope for a given $\Delta n_{\pi}$ pair exhibits strong model dependence.

However, by comparing 3 GeV ultra-peripheral Au–Au collision data with model predictions of the slope values for multiple different $\Delta n_{\pi}$ pairs in Fig.\ref{fig:slope_comparison_band}, we can assess which model more accurately reflects the true neutron skin thickness. This slope-thickness comparison method reduces model dependence in interpreting collision results.


\sect{Discussion} 
In this study we employ the SMASH transport model to simulate ultra-peripheral Au+Au collisions at $\sqrt{s_{\text{NN}}}=3\ \text{GeV}$ and perform an event-by-event analysis of the resulting data.  
By examining the event-by-event distribution of $\Delta n_{\pi}=n_{\pi^{-}}-n_{\pi^{+}}$, we propose a novel approach for estimating the neutron-skin thickness of the Au nucleus.  
Our results indicate that the mean value $\langle\Delta n_{\pi}\rangle$ increases with neutron-skin thickness.

More significantly, the slope of the straight line connecting specific pairs of $\Delta n_{\pi}$ values—namely $(-1,1)$, $(-1,2)$, $(0,1)$ and $(0,2)$—shows a strong linear correlation with neutron-skin thickness.  
When applied to forthcoming experimental collision data, this method promises significantly tighter constraints on the neutron-skin thickness of gold.

Although prior studies have proposed using the ratio of total $\pi^{-}$ and $\pi^{+}$ production cross-sections, defined as
\[
E = \frac{\sigma(\pi^{-}) - \sigma(\pi^{+})}{\sigma(\pi^{-}) + \sigma(\pi^{+})},
\]
to probe differences in neutron and proton distributions within atomic nuclei \cite{Lombard:1988pj,Tellez-Arenas:1987tls}, this approach relies on integrated cross-sections and does not incorporate event-by-event analysis. We also tested the event-by-event results of
\[
E_{\text{event}} = \frac{n_{\pi^{-}} - n_{\pi^{+}}}{n_{\pi^{-}} + n_{\pi^{+}}},
\]
and the results were similar to those obtained using our previously proposed observable  $\Delta n_{\pi} = n_{\pi^{-}} - n_{\pi^{+}}$. However, it should be noted that $E_{\text{event}}$ in event-by-event analyses suffers from degeneracy issues. For instance, in peripheral collisions dominated by neutron-neutron scattering, both the numerator $n_{\pi^{-}} - n_{\pi^{+}}$ and the denominator $n_{\pi^{-}} + n_{\pi^{+}}$ increase as the overlap region grows.

Unlike conventional analyses that rely on integrated yields and the average $\pi^{-}/\pi^{+}$ ratio, our event-wise strategy overcomes limitations inherent in ultra-peripheral collisions, where individual events may contain very few or even zero pions.  
The introduction of $\Delta n_{\pi}$ as an observable circumvents this issue and demonstrates clear sensitivity to neutron-skin thickness.  
This correlation originates from the initial neutron-proton asymmetry in the Au nucleus: the excess of neutrons leads to a larger number of neutron–neutron (nn) collisions relative to proton–proton (pp) collisions.  
Since nn collisions predominantly produce $\pi^{-}$ while pp collisions favour $\pi^{+}$, a systematic excess of $\pi^{-}$ is observed, which becomes more pronounced with increasing neutron-skin thickness.

Despite the novelty of our approach, several limitations must be acknowledged.  
First, the results exhibit a dependence on the choice of transport model: simulations performed with both SMASH and UrQMD yield significantly different outcomes, highlighting the need for careful model selection and validation.  
Second, the omission of Coulomb interactions in SMASH—due to computational constraints—may contribute to the observed discrepancies, as UrQMD incorporates these effects.  
However, by comparing slope values derived from multiple $\Delta n_{\pi}$ pairs across both models with future experimental data, it may be possible to identify which model more accurately reflects physical reality, thereby mitigating model-dependent uncertainties.

This study introduces an event-wise analysis of $\Delta n_{\pi}$ distributions as a new tool for probing neutron-skin thickness, offering both methodological advances for nuclear physics and valuable theoretical insights for astrophysics.

\sect{Summary}
In summary, we have explored the potential of the event-by-event distribution of the pion-yield difference $\Delta n_{\pi}=n_{\pi^{-}}-n_{\pi^{+}}$ as a novel observable for constraining the neutron-skin thickness $\Delta r_{np}$ of the Au nucleus.  
Using the SMASH model, we simulated ultra-peripheral Au+Au collisions at $\sqrt{s_{NN}}=3$ GeV.  

Our analysis yields four key results: the mean value of $\langle\Delta n_{\pi}\rangle$, and the Pearson correlation and mutual information of $(\pi^{-}+\pi^{+})$ and $(\pi^{-}-\pi^{+})$ all increase with growing $\Delta r_{np}$. More importantly, the slopes between specific $\Delta n_{\pi}$ pairs in the distribution exhibit a strong linear dependence on the neutron-skin thickness.  
Among the pairs examined, $\Delta n_{\pi}=(-1,1)$, $(-1,2)$, $(0,1)$ and $(0,2)$ showed the highest sensitivity.

This approach offers a powerful new method to precisely determine the neutron-skin thickness of gold nuclei, which is essential for constraining the nuclear symmetry energy and the equation of state of asymmetric nuclear matter.  
A comparison with UrQMD simulations indicated a non-negligible model dependence, possibly due to the absence of Coulomb interactions in the SMASH setup.  
To address this uncertainty, we propose leveraging slopes from multiple $\Delta n_{\pi}$ pairs in conjunction with future experimental data, which may also help identify the most suitable transport model.  


\begin{acknowledgments}
This work was supported in part by the NSFC under grant No.~12435009 and 12075098. We acknowledge helpful discussions with Xin Dong and Xiao-Feng Luo. We gratefully acknowledge the extensive computing resources provided by the nuclear computing center of Central China Normal University.
\end{acknowledgments}

\bibliography{ref}

\end{document}